\documentclass[colorlinks]{goose-article}

\usepackage{booktabs}
\usepackage[toc,page,titletoc]{appendix}

\crefname{supp}{Supplement}{Supplements}

\title{Dynamical heterogeneities of thermal creep in pinned interfaces}

\author[1]{Tom~W.J.~de~Geus}
\author[2]{Alberto~Rosso}
\author[1]{Matthieu~Wyart}

\affil[1]{Institute of Physics, \'{E}cole Polytechnique F\'{e}d\'{e}rale de Lausanne (EPFL), Switzerland}

\affil[2]{LPTMS, CNRS, Univ.\ Paris-Sud, Universit\'{e} Paris-Saclay, 91405 Orsay, France}

\hypersetup{pdfauthor={T.W.J. de Geus}}

\begin{document}

\twocolumn[
    \begin{@twocolumnfalse}

        \maketitle

        \begin{abstract}
            \noindent
            Disordered systems under applied loading display slow creep flows at finite temperature, which can lead to the material rupture.
            Renormalization group arguments predicted that creep proceeds via thermal avalanches of activated events.
            Recently, thermal avalanches were argued to control the dynamics of liquids near their glass transition.
            Both theoretical approaches are markedly different.
            Here we provide a scaling description that seeks to unify dynamical heterogeneities in both phenomena, confirm it in simple models of pinned elastic interfaces, and discuss its experimental implications.
        \end{abstract}

    \end{@twocolumnfalse}
]
\sloppy

\paragraph{Introduction}

The physical properties of a material are often controlled by the interfaces embedded in it.
This is the case for dislocations in crystalline materials \cite{Miguel2001,Madec2003,Zaiser2006}, crack fronts in fracture \cite{Bouchaud1997}, frictional interfaces \cite{Scholz1998}, or domain walls in magnets \cite{Zapperi1998}.
Pinned by impurities, they dramatically increase the material's strength, toughness, or hysteresis.
At the critical force (or stress, or magnetic field) $f_c$, the interfaces undergo a \emph{depinning transition} \cite{Kardar1998,Fisher1998} and move via large reorganizations, called avalanches.
Their maximal linear extent diverges approaching the critical point, $\ell_{\text{dep}} (f) \sim |f-f_c|^{-\nu}$ and their size is power law distributed, $P(S) \sim S^{-\tau}$.
Moreover, the interface shape is a self-similar shape with a roughness exponent $\zeta$.
These exponents $\tau, \zeta,\nu,\ldots$ are now well understood and related by scaling relations summarized in \cref{tab:depinning}.

The dynamics below $f_c$, in the so-called {\em creep regime}, is more controversial.
Here, the motion is controlled by the thermal activation over the pinning potential \cite{Ferrero2021}.
Creep is fundamental for a variety of phenomena, ranging from the failure of materials \cite{Leocmach2014,Divoux2011,Siebenburger2012,Rosti2010} to the rupture of frictional interfaces and faults \cite{Baumberger1994,Bureau2001,Bar-Sinai2013,Harris2017,Vincent-Dospital2020}.
In \cite{Chauve2000}, a study of pinned interfaces based on the functional renormalization group (FRG), predicted the existence of two distinct length scales:

(i) The first length, $\ell_{\text{opt}}(f)$, corresponds to the extent of the elementary excitations.
Elementary excitations correspond to the minimal motion in the direction of the applied force that brings the interface to a new local minimum with smaller energy.
They can be treated as being irreversible, since the probability of going backward once an excitation is triggered is small.
$\ell_{\text{opt}}(f)$ grows as $\ell_{\text{opt}}(f) \sim f^{-\nu_{\text{eq}}}$ when the force $f$ vanishes.
Below this scale, the interface is at equilibrium with a roughness exponent~$\zeta_{\text{eq}}$ \cite{Ioffe1987,Nattermann1987,Agoritsas2012}.

(ii) The second, much larger length, $\ell_c(T,f)$, identifies with the cut-off of the {\em thermal avalanches} induced by the interaction between elementary excitations.
Below this length, the roughness exponent is $\zeta$ and avalanches are depinning-like.
This cut-off also diverges, as
\begin{equation}
    \ell_c(T,f) \sim T^{-\sigma} f^{-\lambda}.
\end{equation}
The FRG equations cannot be solved exactly.
In \cite{Chauve2000} an ansatz was proposed that leads to $\sigma = \nu/\beta$ and $\lambda = \nu_{\text{eq}}+\mu_{\text{eq}} \nu/\beta $ \footnote{
    The depinning exponent $\beta$ controls, at zero temperature, the velocity of the interface, which vanishes as $v\sim (f-f_c)^\beta$ at $f_c$.
    The equilibrium exponent $\mu_{\text{eq}}$ appears in the creep formula $\ln v\sim f^{- \mu_{\text{eq}}}$.
    This behaviour has been confirmed experimentally in \cite{Lemerle1998}.
}.

Efficient optimization algorithms at vanishing temperature $T=0^+$ have studied the case of a one-dimensional domain wall moving in a medium in two dimensions \cite{Kolton2006,Kolton2009,Ferrero2017}.
They confirmed the scaling of $\ell_{\text{opt}}(f)$, the crossover of the roughness exponents from equilibrium to depinning, as well as the presence of thermal depinning avalanches.
Experiments on a one-dimensional domain wall in a ferromagnetic film showed the validity of the creep formula for the wall velocity, which is an indirect check for the scaling of $\ell_{\text{opt}}(f)$ \cite{Lemerle1998,Metaxas2007}.
Recently, depinning thermal avalanches have been observed via Magneto-optic Kerr effect (MOKE) techniques \cite{Grassi2018,Durin2023} and in crumpled paper \cite{Lahini2023,Shohat2023}.
However, the determination of $\ell_c(T,f)$ remains an open problem -- finite temperature simulations using Langevin dynamics were inconclusive \cite{Kolton2009}.

Recently, it was proposed that the concept of thermal avalanches extends beyond driven systems, and explains the dynamical heterogeneities observed in supercooled liquids \cite{Tahaei2023}.
In that case, it was argued that $\ell_c(T) \sim T^{-\sigma}$ with $\sigma\geq(1+\theta)/d$, whereby $\theta$ is an exponent that characterizes the distribution of excitations \cite{Lin2014a,Lin2016}.
In the context of the depinning transition $\theta = 0$.
In two dimensions, this inequality was found to be saturated in mesoscopic models of liquids \cite{Tahaei2023}.
Arguments in \cite{Tahaei2023} should also go through in the depinning context, yet the bound $\sigma\geq 1/d$ is much below the FRG predictions.

In this Letter, we propose a scaling theory of dynamical heterogeneities for both phenomena.
We introduce a stylized model for creep of depinning interfaces, where the optimized elementary excitations are coarse-grained at the scale of a single block.
At zero temperature, the model reduces to the cellular automata used to study the depinning transition and its exponents.
At finite temperature, a similar model was discussed in \cite{Vandembroucq2004,Purrello2017}, but not used to study the creep regime.
We provide theoretical arguments indicating that
\begin{equation}
    \sigma = \nu
    \label{eq:result1}
\end{equation}
both for depinning and for the glass transition (see below for a definition of $\nu$ in the latter case).
Moreover, we argue that in the depinning case, the relationship between the age $t$ of an avalanche and its spatial extent follows, as for liquids \cite{Tahaei2023}:
\begin{equation}
    \ell_c(t) \sim T^{-\nu} \left[ \ln (\tau_\alpha / t) \right]^{-\nu}
    \label{eq:result2}
\end{equation}
where $\tau_\alpha$ is a characteristic timescale at which a fraction of the interface has moved.
We find that all our predictions hold quantitatively in a mesoscopic model of depinning.

\begin{table}[htp]
    \centering
    \caption{
        Critical exponents of the depinning transition of an elastic manifold with short-range interactions in $d$ dimensions, embedded in $d + 1$ dimensions.
        Exponent values correspond to $d=2$.
    }
    \label{tab:depinning}
    \begin{tabular}{llll}
        \toprule
        Name    & Expression                     & Scaling               & Value   \\
        \midrule
        $\tau$  & $P(S) \sim S^{-\tau}$          & $2 - 2 / (d + \zeta)$ & $1.27$  \\
        $\zeta$ & $S \sim \ell^{d + \zeta}$      &                       & $0.753$ \\
        $\nu$   & $\ell_c \sim |f_c - f|^{-\nu}$ & $1 / (2 - \zeta)$     & $0.80$  \\
        \bottomrule
    \end{tabular}
\end{table}

\begin{figure}[htp]
    \subfloat{\label{fig:extremal:pdf_x}}
    \subfloat{\label{fig:extremal:pdf_s}}
    \subfloat{\label{fig:extremal:ellc}}
    \centering
    \includegraphics[width=\linewidth]{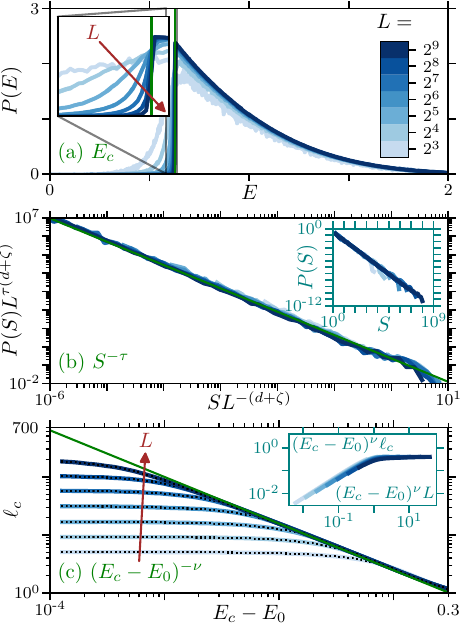}
    \caption{
        At $T = 0^+$:
        \protect\subref{fig:extremal:pdf_x}
        Distribution of activation barriers
        $P(E)$ at random snapshots, for different system sizes $L^d$ as specified by the color bar.
        As $L\rightarrow \infty$, $P(E)$ displays a gap at $E_c\approx 0.6284 \pm 0.0001$.
        (\protect\subref*{fig:extremal:pdf_s}, \protect\subref*{fig:extremal:ellc}) Failure events are segmented into avalanches as sequences for which the activation barrier of the failing block $E_{\min} < E_0$.
        \protect\subref{fig:extremal:pdf_s}
        Inset: avalanches distribution $P(S)$ for $E_0 = E_c$.
        The prediction
        $P(S) \sim S^{-\tau} f(S/L^{d + \zeta})$ where $f$ is a scaling function, is tested in the main panel by rescaling the axis: a collapse is apparent with the depinning exponents of \cref{tab:depinning}.
        \protect\subref{fig:extremal:ellc}
        Main panel: linear extent of avalanches $\ell_c$ as a function of the threshold $E_0 < E_c$, for different system size $L$.
        The scaling behavior, $\ell_c = (E_c - E_0)^{-\nu} g((E_c - E_0)^{-\nu}/L)$ is confirmed in the inset for $\nu = 1 / (2 - \zeta) \approx 0.80$.
    }
    \label{fig:extremal}
\end{figure}

\paragraph{Model}
We model an elastic interface in a mesoscopic manner, coarse-grained on the excitation scale $\ell_{\text{opt}}(f)$.
Such a model is the basis of both our theoretical arguments, and of our numerical study below.
The system is a lattice of $L^d$ blocks.
The block $i$ is characterized by an applied force $f_i$ (the sum of the driving force $f$ and elastic forces coming from neighboring blocks), and by a yielding threshold $f_i^y$.
Its stability is governed by the distance to yielding $x_i \equiv |f_i| - f_i^y$, corresponding to an activation barrier $E_i\equiv |x_i|^\alpha \, \mathrm{sign}(x_i)$ with some $\alpha>0$.
Our arguments below do not depend on the value of $\alpha$.

At a finite temperature $T > 0$ (that we express in the units of the Boltzmann constant $k_B$), the time to failure $\tau$ of a block is exponentially distributed $P(\tau) = \exp(-\tau / \tau_i) / \tau_i$.
Here the characteristic time is $\tau_i = \tau_0 \exp(E_i / T)$ if $E_i \geq 0$ and $\tau_i = \tau_0$ otherwise, where $\tau_0$ is a fixed microscopic timescale.
After failure, the force $f_i$ drops to a small random value, and a fresh value of the yield force $f_i^y$ is sampled from a fixed distribution.
The force drop is redistributed equally to the nearest neighboring blocks, as follows from the presence of elastic interactions.
Finally, the position of the interface at block $i$, $u_i$, is increased by the value of the force drop (using an elastic constant of one).

It is well known that such a cellular automaton has a threshold force $f_c$, such that if the applied force $f = \sum_i f_i / L^d$ is larger than a depinning threshold $f_c$, the dynamics does not stop even if $T=0$.
In our arguments we consider that $f<f_c$ and $T>0$, while numerics are performed with $f=f_c/2$, $d=2$ and $\alpha=3/2$.
We express forces in units of average yield threshold $\bar{f}^y$, time in units of $\tau_0$, temperature in units of $k_B (\bar{f}^y)^\alpha$, and length in units of the lattice spacing.
Further numerical details are specified in \cref{sec:si:model}.

\paragraph{Extremal dynamics}
For finite $L$, if $T$ is vanishingly small, the site with the smallest activation barrier $E_{\min} \equiv \min_i(E_i)$ always relaxes first.
It is an example of {\it extremal dynamics} noted here as $T=0^+$.
It is well studied in the context of self-organized criticality~\cite{Paczuski1996,Vandembroucq2004,Purrello2017}.
As $L\rightarrow \infty$, the distribution of energy barriers $P(E)=\sum_i\delta(E-E_i)/L^d$ must have a compact support, and be zero below a force-dependent value $E_c$ \footnote{
    The probability to trigger to instability a block with $E_i>E_c$ is vanishing as $L\rightarrow\infty$.
    Because the number of blocks is extensive in the energy range $[E_c,E_i]$, $E_i>0$ can never be the extremal, minimum, energy among all blocks.
}.
$E_c$ must grow as the applied force decreases, and vanishes at the depinning threshold $f=f_c$.
Thus in the case $f<f_c$ considered here one has $E_c>0$.

For finite $L$, $E_c$ in not sharply defined but displays sample to sample fluctuations illustrated in \cref{fig:extremal:pdf_x}.
After averaging the distribution of $E_c$ over various configurations, $P_L(E)$ has a standard deviation $\Delta E_c$ that decays algebraically with $L$ as
\begin{equation}
    \Delta E_c\sim L^{-1/\nu}.
    \label{eq:nu}
\end{equation}
Note that for elastoplastic models used to study the glass transition \cite{Tahaei2023,Ozawa2023}, \cref{eq:nu} defines $\nu$, which will take a different value than for the elastic interface.

For extremal dynamics, avalanches can be defined as a sequence of $S$ fast events where the most unstable site $E_{\min}$ always satisfies $E_{\min} < E_0$, where $E_0$ is some chosen threshold ~\cite{Paczuski1996,Purrello2017}.
In the thermodynamic limit, it is clear that the sequence never stops if $E_0 > E_c$.
It is critical at $E_0 = E_c$, and falls in the depinning universality class with $P(S) \sim S^{-\tau}$, with a cut-off $S_c \sim L^{d + \zeta}$ at finite $L$.
It is confirmed in \cref{fig:extremal:pdf_s} with exponents indeed given by \cref{tab:depinning} (see \cref{sec:si:clusters} for an illustration on the method and \cref{sec:si:zeta} for further supporting measurements).

\begin{figure}[htp]
    \centering
    \includegraphics[width=\linewidth]{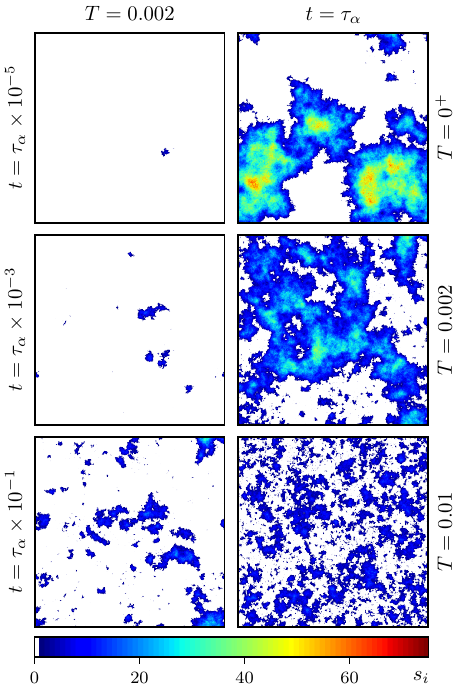}
    \caption{
        Left column: number of times $s_i$ that a block yielded during a time interval $t$ (increasing from top to bottom) for our lowest temperature $T = 0.002$ and our biggest system $L = 2^9 = 512$.
        Right column: same quantity for different temperatures (increasing from top to bottom) at $t = \tau_\alpha$, defined as the time for which half of the sites yielded at least once.
    }
    \label{fig:snapshots}
\end{figure}
For $E_0<E_c$, avalanches will be cut off once they reach some spatial extent $\ell_c$.
$\ell_c$ must be the length scale where the local fluctuations of the gap $\Delta E_c(\ell_c)$ are of order of the difference $E_c-E_0$: clearly, on a smaller scale, this difference is irrelevant.
Using \cref{eq:nu} yields:
\begin{equation}
    \ell_c\sim (E_c-E_0)^{-\nu}.
    \label{eq:result3}
\end{equation}
This result is confirmed numerically in \cref{fig:extremal:ellc}.
Here, we define the linear extent of an avalanche as $\ell \equiv A^{1 / d}$, where the `area' $A$ is the number of sites that yielded at least once.
The cut-off $\ell_c$ is then defined as the ratio of moments $\ell_c \equiv \langle \ell^2 \rangle / \langle \ell \rangle$ where the average $\langle \dots \rangle$ is made over avalanches.
We define $E_c$ to provide the best collapse of \cref{eq:result3} in the inset of \cref{fig:extremal:ellc}.

\paragraph{Finite temperature}

\begin{figure}[htp]
    \subfloat{\label{fig:thermal:time}}
    \subfloat{\label{fig:thermal:time_collapse}}
    \centering
    \includegraphics[width=\linewidth]{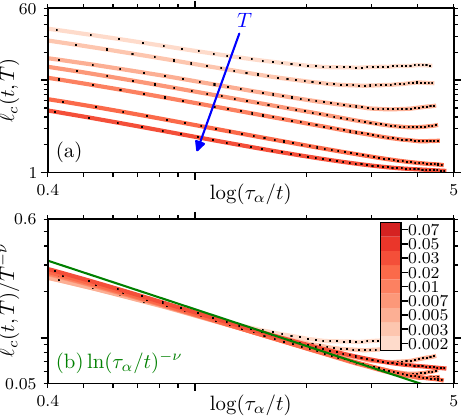}
    \caption{
        \protect\subref{fig:thermal:time}
        Extent of connected clusters $\ell_c(t)$ after some time interval $t$, for different temperatures, as indicated in the color bar.
        \protect\subref{fig:thermal:time_collapse} The prediction $\ell_c(t) \sim T^{-\nu} \left[ \ln (\tau_\alpha / t) \right]^{-\nu}$ is tested by rescaling the axis, which indeed approximately collapses the different curves.
        The straight green line indicates the prediction $\ln (\tau_\alpha / t)^{-\nu}$.
    }
    \label{fig:thermal}
\end{figure}

At a finite but small enough (to be specified below) temperature $T$ and fixed $L$, the above picture remains true, but the dynamics now occur on a finite time scale.
The duration $t$ of an avalanche defined with threshold $E_0$ will be of order of the longest waiting time between events within it, of order $\exp(E_0 / T)$, thus
$t\sim \exp(E_0 / T)$.
The largest avalanches filling up the system occur on some time $\tau_\alpha\sim \exp(E_c/T)$.
We extract numerically $\tau_\alpha$ as the time at which half of the system fails at least once, which indeed is proportional to $\exp(E_c/T)$ as we test in \cref{sec:si:tau_alpha}.
Injecting these definitions in \cref{eq:result3}, one gets for $t \leq \tau_\alpha$:
\begin{equation}
    \label{eq:ellc}
    \ell_c (t, T) \sim T^{-\nu} \left[ \ln (\tau_\alpha / t) \right]^{-\nu}.
\end{equation}
A similar argument was proposed for supercooled liquids, and matched the growth of dynamical heterogeneities with time in molecular dynamics observations well \cite{Tahaei2023}.
This prediction can be tested in the depinning case by considering the pattern of relaxations occurring in some time interval $t$, as illustrated in the left panel of \cref{fig:snapshots}.
We extract avalanches from these figures as connected clusters, and $\ell_c(t,T) \equiv \langle \ell^2(t) \rangle_T / \langle \ell(t) \rangle_T$.
Here the average $\langle \ldots \rangle_T$ is made on all clusters observed at temperature $T$ in intervals of duration $t$, and the extent $\ell$ is defined as above (the square root of the cluster area in our $d = 2$ numerics).
Our prediction is confirmed in \cref{fig:thermal} (see \cref{sec:si:clusters} for illustration of the method, and \cref{sec:si:thermal_avalanches} for further supporting measurements on both the avalanche exponents as well as our prediction in \cref{eq:ellc}).

We now focus on our central question: the correlation length $\ell_c(T)$ in an infinite system.
The right panel of \cref{fig:snapshots} shows the interface motion during a time interval of $\tau_\alpha$ (where half of the sites relaxed), for different temperatures.
Clearly, the dynamics is more and more correlated under cooling.
To explain these observations, we perform a finite size scaling argument.
Consider an infinite system cuts into subsystems of size $L$.
The fluctuations $\Delta E_c$ of the gaps in different subsystems follows $\Delta E_c(L) \sim L^{-1/\nu}$.
If $T\gg \Delta E_c$, these subsystems relax essentially at the same speed, while if $T \ll \Delta E_c(L)$ their pace is very different.
Thus, dynamical heterogeneities should exist up to a length $\ell_c$ such $T\sim \Delta E_c(\ell_c)$, implying:
\begin{equation}
    \label{eq:xi}
    \ell_c(T) \sim T^{-\nu}.
\end{equation}
Note that this result also identifies $\ell_c(T)$ with the avalanche extent occurring on a timescale $\cal{O}(\tau_\alpha)$, according to \cref{eq:ellc}.

In the case of elastic interfaces, dynamical correlations also affect the structural correlations in the interface geometry.
In particular, we expect the depinning roughness exponent $\zeta$ to hold up to a scale $\ell_c(T) \sim T^{-\nu}$.
Above this length scale, avalanches become independent and they are randomly triggered.
As in the Edwards-Wilkinson model, the roughness becomes $\zeta_{EW} =(2-d)/2$ for $d<2$ and logarithmic for $d=2$.

\begin{figure}[htp]
    \subfloat{\label{fig:thermal:structure}}
    \subfloat{\label{fig:thermal:structure_collapse}}
    \centering
    \includegraphics[width=\linewidth]{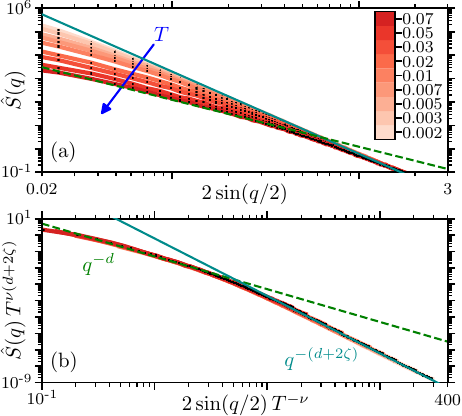}
    \caption{
        \protect\subref{fig:thermal:structure}
        Structure factor of the position of the interface $u(x)$ at location $x$ (see main text), for different temperatures (see color bar).
        All results for our largest system size $L = 2^9$.
        At small distances (large $q$) we find a fast decay, $\sim 1/q^{d+2 \zeta}$, consistent with the depinning exponent $\zeta$ in \cref{tab:depinning} (solid blue line).
        At large distances (small $q$) we observe a slower decay $\sim 1/q^{d}$ (dashed green line) consistent with a logarithmic roughness.
        \protect\subref{fig:thermal:structure_collapse} The collapse of data indicates that the crossover scale is $\ell_c(T) \sim T^{-\nu}$.
    }
    \label{fig:thermal_structure}
\end{figure}

This prediction is verified by considering the structure factor $\hat{S}(q) \equiv \langle \hat{u}(\vec{q}) \hat{u}({-\vec{q}}) \rangle$, with $u(\vec{x})$ the position of the interface at location $\vec{x}$; $\hat{u}(\vec{q})$ is its Fourier transform and $q$ the Euclidean norm of wave vector $\vec{q}$.
The average $\langle \ldots \rangle$ is taken on independent snapshots taken at random times.
For a self-affine interface with roughness exponent $\zeta$, the structure factor scales as $\hat{S}(q) \sim q^{-(d + 2 \zeta)}$ (e.g.~\cite{Ramasco2000,Schmittbuhl1995a}) and as $\hat{S}(q) \sim q^{-d}$ for a logarithmic roughness.
The crossover between the two regimes occurs at a scale $q_c(T) \sim 1/\ell_c(T)$.
We verify such scaling in \cref{fig:thermal:structure} and find that the crossover scale from \cref{eq:xi} indeed collapses data for different $T$ in \cref{fig:thermal:structure_collapse}.

\paragraph{Conclusion}
We have provided a description of thermal avalanches which may characterize dynamical heterogeneities in both the creep flows of pinned interfaces and in supercooled liquids.
As proposed for liquids \cite{Tahaei2023}, we argued in the context of depinning that the growth of avalanche size with time is logarithmic, and tested this prediction in a simple model.
This prediction could be tested in experiments monitoring the domain wall dynamics \cite{Grassi2018,Durin2023}
or by the motion of dislocations in crystalline materials \cite{Rodney2023}.

Our central prediction concerns the correlation length $\ell_c(T)$.
The scaling of the correlation length $\ell_c(T) \sim T^{-\nu}$ differs from the FRG prediction of \cite{Chauve2000}.
The latter was obtained from an Ansatz solution, it will be interesting to see if our prediction can be derived from the original FRG equations.
It will also be important to verify \cref{eq:xi} from direct simulations of a continuous elastic line at finite temperature, or in experiments by recording the dynamics at different temperatures using a cryostat, as in \cite{Grassi2018}.

For supercooled liquids, we have argued that $\ell_c(T) \sim T^{-\nu}$, which should be, in general, distinct from the bound $(1 + \theta)/d$ proposed in \cite{Tahaei2023}.
In two dimensions, an elastoplastic model of the glass transition predicted $(1+\theta)/d\approx 0.8$ and $\nu\approx 0.9$ \footnote{$\nu$ is noted $1/{\tilde{d_f}\tilde\sigma}$ in that reference, where it is extracted from \cref{eq:result3}.}, and both predictions are consistent with the observation $\ell_c(T)\sim T^{-0.8}$.
We expect the similarity between these exponents to be a coincidence, which for example does not occur for the pinned interface investigated here.
We expect this similarity to disappear in the practically relevant case of three-dimensional liquids.
This point can now be tested in three-dimensional elastoplastic models, by extracting both exponents $\nu$ and $\theta$.
Measuring $\nu$ would lead to quantitative prediction for the temporal or thermal behavior of dynamical heterogeneities that could be tested in experiments \cite{Weeks2000}.

\paragraph{Acknowledgements}
E.~El Sergany is thanked for exploratory numerical tests, and L.~Berthier, G.~Biroli, C.~Brito, C.~Gavazzoni, D.~Korchinski, W.~Ji, Y.~Lahini, M.~M\"{u}ller, M.~Ozawa, M.~Pica Ciamarra, M.~Popovi\'{c}, D.~Shoat, and A.~Tahei for discussions.
T.G.~acknowledges support from the Swiss National Science Foundation (SNSF) by the SNSF Ambizione Grant PZ00P2{\_}185843.
M.W.~acknowledges support from the Simons Foundation Grant (No.~454953 Matthieu Wyart).

\bibliographystyle{unsrtnat}
\bibliography{library}

\clearpage
\appendixpageoff
\appendixtitleoff
\renewcommand{\appendixtocname}{Supplementary material}

\begin{appendices}

    \crefalias{section}{supp}
    \twocolumn[
        \begin{center}
            \Large
            \textbf{Supplementary material}
            \normalsize
        \end{center}
    ]

    \renewcommand\theequation{S\arabic{equation}}
    \setcounter{equation}{0}

    \renewcommand\thefigure{S\arabic{figure}}
    \setcounter{figure}{0}

    \section{Model details}
    \label{sec:si:model}

    We list the remaining details of our model.

    \paragraph*{Interactions}

\begin{figure}[htp]
        \centering
        \includegraphics[width=0.65\linewidth]{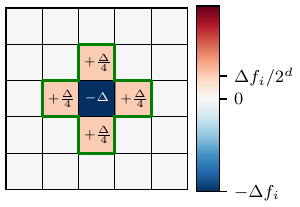}
        \caption{
            Stress redistribution among the nearest neighbors (outlined in green) of a failing block in $d = 2$ dimensions.
        }
        \label{fig:model}
    \end{figure}

    The elastic interactions are such that if a block $i$ fails, it loses its force up to a small random number, $\epsilon$, such that $f_i(t + \tau_i) = f_i(t) - \Delta f_i$ with $\Delta f_i = f_i(t) - \epsilon$.
    Thereby, $\epsilon$ is a random number distributed normally as $\mathcal{N}(0, 0.01)$, whereby the notation is such that its mean in $0$ and its standard deviation is $0.01$.
    The position of the interface $u_i(t + \tau_i) = u_i(t + \tau_i) + \Delta f_i$ (using an elastic constant of one).
    The force drop $\Delta f_i$ of the failing block $i$ is entirely redistributed to the nearest neighbors, $j$, as shown in \cref{fig:model} for our case in $d = 2$ dimensions.
    Consequently, the applied force $f = \sum_i f_i / L^d$ is constant at all times.

    \paragraph*{Forces}

    We take $f = 0.3 \approx f_c / 2$.
    We estimated $f_c$ by driving the interface with a weak spring.
    Furthermore, the yield threshold $f_i^y$ of each block $i$ is random according to a normal distribution $f_i^y = \mathcal{N}(1, 0.3)$, truncated at $0$.

    \paragraph*{Preparation}

    The force in each block is initialized random according to a normal distribution, $f_i = \mathcal{N}(f, 0.1)$, which is convoluted once with the elastic kernel from \cref{fig:model}.
    Any unstable block $x_i = |f_i| - f_i^y < 0$ is then failed, and the stress is redistributed to its nearest neighbors.
    This is repeated until all blocks are stable.
    The system is then driven for at least $100 L^d$ steps at the temperature used for the measurement (or using ``extremal dynamics'' if $T = 0^+$).

    \section{Roughness exponents}
    \label{sec:si:zeta}

\begin{figure}[htp]
        \centering
        \includegraphics[width=\linewidth]{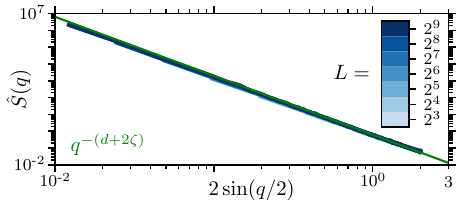}
        \caption{
            At $T = 0+$, the structure factor $\hat{S}(q) \equiv \langle \hat{u}(\vec{q}) \hat{u}({-\vec{q}}) \rangle$, with $u(\vec{x})$ the position of the interface at location $\vec{x}$ at random snapshots (we consider an ensemble at random times in the steady state, for random realizations).
            $\hat{u}(\vec{q})$ is the Fourier transform of the position of the interface, with $\vec{q}$ the wave vector and $q$ its Euclidean norm.
            In practice, we plot simply the subset $q_x > 0$ and $q_y = 0$ (whereby we work in $d = 2$ dimensions).
            The interface is rough such that $\delta u \sim r^\zeta$ with $r$ the Euclidean distance, corresponding to $\hat{S}(q) \sim q^{-(d + 2 \zeta)}$ as shown in green for the depinning prediction in \cref{tab:depinning}.
        }
        \label{fig:extremal:structure}
    \end{figure}

\begin{figure}[htp]
        \centering
        \includegraphics[width=\linewidth]{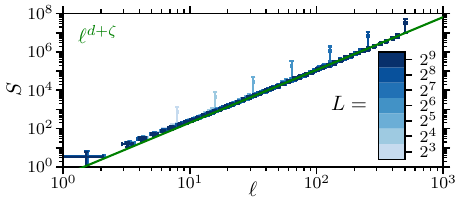}
        \caption{
            Fractal dimension of avalanches at $T = 0^+$.
            Avalanches are defined as a sequence for which the activation barrier of the failing block $E_{\min} < E_0$.
            We take $E_0 = E_c$ and show $S$, the total number of fails in the sequence, and $\ell^d$, the number of unique blocks in the sequence.
            The scaling, in green, $S \sim \ell^{d + \zeta}$ is consistent with the depinning prediction in \cref{tab:depinning}.
        }
        \label{fig:extremal:fractal}
    \end{figure}

    The geometry of avalanches is consistent with a roughness exponent $\zeta$, with its value predicted by the depinning theory recalled in \cref{tab:depinning}.
    We support this by two measurements at $T = 0^+$.
    First, we measure the roughness of the interface at random snapshots in the steady state.
    In particular, we measure the structure factor and find that it is consistent with $\hat{S}(q) \sim q^{-(d + 2 \zeta)}$ (see main text), as shown in \cref{fig:extremal:structure} (see e.g.~\cite{Ramasco2000,Schmittbuhl1995a} for discussions on the use of the structure factor to measure the roughness exponent of a self-affine interface).
    Second, we measure the fractal dimension of avalanches.
    We define avalanches as a sequence for which the activation barrier $E_{\min} < E_0$ with $E_0 = E_c$.
    We plot the total number of fails in each sequence, $S$ (the `duration' of each sequence), as a function of spatial extent $\ell$ (with $\ell^d$ the number of unique blocks in the sequence) in \cref{fig:extremal:fractal}.
    We find that it is consistent with $S \sim \ell^{d + \zeta}$.

    \section{Avalanche identification}
    \label{sec:si:clusters}

    \paragraph{Extremal dynamics}

\begin{figure}[htp]
        \centering
        \includegraphics[width=\linewidth]{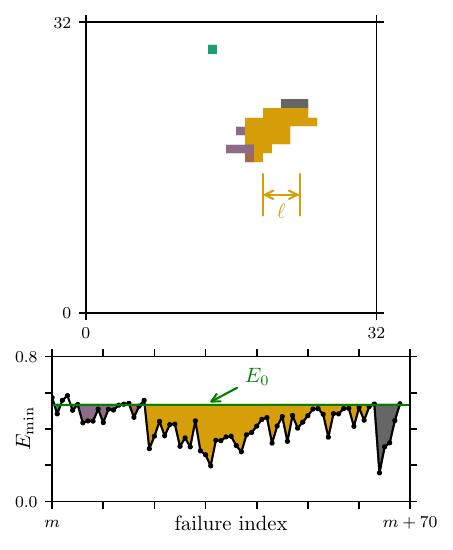}
        \caption{
            Identification of avalanches for extremal dynamics ($T = 0^+$).
            An avalanche is identified as a connected sequence of events for which the activation barrier of the failing block $E_{\min} < E_0$ for some threshold $E_0$.
            An example is shown for a small subset of failures in a system of size $L = 2^5 = 32$ with $E_0 = 0.85 E_c$.
            The bottom panel shows $E_{\min}$ as a function of the number of failures (starting from an arbitrary point $m$ in the sequence).
            The identified avalanches based on the threshold $E_0$ are indicated by different colors.
            The spatial organization of these five avalanches is shown in the top panel.
            The linear extent of the largest avalanche is also indicated, using $\ell$.
        }
        \label{fig:extremal_method}
    \end{figure}

    We measure a sequence of failures starting from a random snapshot somewhere in the steady state (for which each block already yielded many times since preparation).
    We record the activation barrier $E_{\min}$ and the spatial location of each failure.
    We define avalanches as sequences of failures for which $E_{\min} < E_0$, with $E_0$ some threshold.
    We illustrate this in \cref{fig:extremal_method} for a small subset of failures in a small system.
    These avalanches are characterized by $S$, the total number of failures in sequence, and $A$ the number of blocks that had at least one failure since the beginning of the avalanche.
    Then, $\ell \equiv A^{1/d}$ is used as a proxy of the linear extent of the avalanche.

    \paragraph{Finite temperature}

\begin{figure}[htp]
        \subfloat{\label{fig:snapshot:slip}}
        \subfloat{\label{fig:snapshot:s}}
        \centering
        \includegraphics[width=\linewidth]{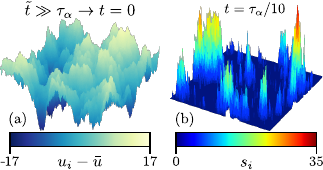}
        \caption{
            \protect\subref{fig:snapshot:slip}
            Snapshot of the position $u_i$ of the interface at a random physical time $\tilde{t}$ in the steady state.
            The average $\bar{u} = \sum_i u_i / L^d$ has been subtracted, whereby we note that the values of both $\tilde{t}$ and $\bar{u}$ are irrelevant.
            \protect\subref{fig:snapshot:s}
            Avalanches are then identified by taking the interface entirely flat at $\tilde{t}$ (setting the duration $t = 0$) and recording the number of times $s_i$ that each block $i$ fails at different durations $t$.
            As an example, we show $s_i$ at $t = \tau_\alpha / 10$ (with $\tau_\alpha$ the duration at which half of the blocks yielded at least once).
            This example is taken for our lowest temperature $T = 0.002$ and our biggest system size $L = 2^9$.
        }
        \label{fig:snapshots_method}
    \end{figure}

    We again measure avalanches starting from an ensemble of independent snapshots at random times in the steady state.
    As an example, we show a snapshot of the position of the interface at a random time in the steady state in \cref{fig:snapshot:slip}.
    Starting from this snapshot which marks a duration $t = 0$ we record the number of times $s_i$ that each block $i$ fails at different times $t$ (see example in \cref{fig:snapshot:s}).

\begin{figure}[htp]
        \subfloat{\label{fig:clusters:projection}}
        \subfloat{\label{fig:clusters:clusters}}
        \centering
        \includegraphics[width=\linewidth]{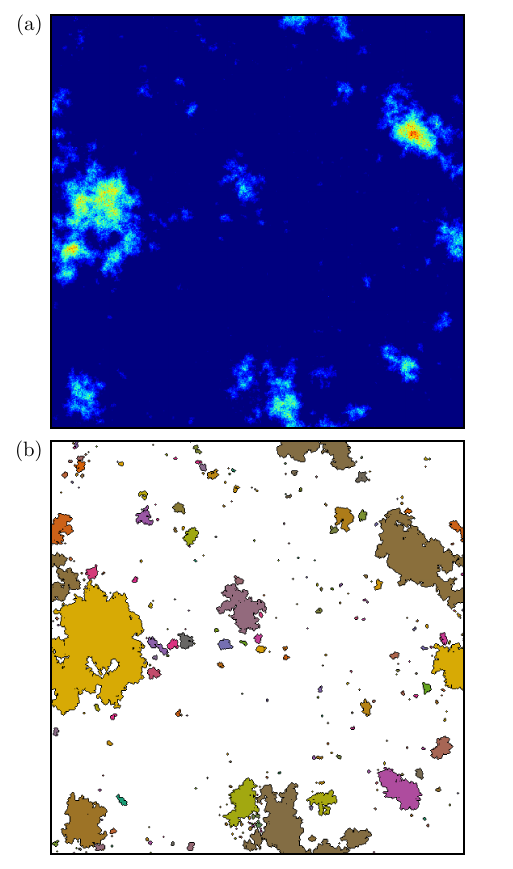}
        \caption{
            \protect\subref{fig:clusters:projection} Projection of \cref{fig:snapshots_method} with the same color bar.
            \protect\subref{fig:clusters:clusters} Identified clusters (colors with black outline), see text for details.
        }
        \label{fig:snapshots_clusters}
    \end{figure}

    Individual avalanches are then identified as connected clusters of blocks that failed at least once.
    An example is shown in \cref{fig:snapshots_clusters}.
    \cref{fig:clusters:projection} shows a projection of \cref{fig:snapshots_method} with the same color bar.
    The identified clusters are shown in \cref{fig:clusters:clusters}, whereby each cluster is assigned a different color, and a black outline is drawn around each cluster.
    The connectedness of clusters is decided using the same nearest-neighbor kernel as used for the elastic interactions, see \cref{fig:model}.
    Once identified, we extract for each cluster (``avalanche'') the number of blocks $A$ in that cluster (resulting in $\ell \equiv A^{1 / d}$), and the total number of fails $S$ of all blocks in that cluster.
    Note that the clusters are geometrically identified at each time $t$, i.e.~their identification is independent of previous times.

    \section{Avalanche statistics}
    \label{sec:si:thermal_avalanches}

\begin{figure}[htp]
        \subfloat{\label{fig:thermal:fractal}}
        \subfloat{\label{fig:thermal:pdf_s}}
        \subfloat{\label{fig:thermal:pdf_s_collapse}}
        \centering
        \includegraphics[width=\linewidth]{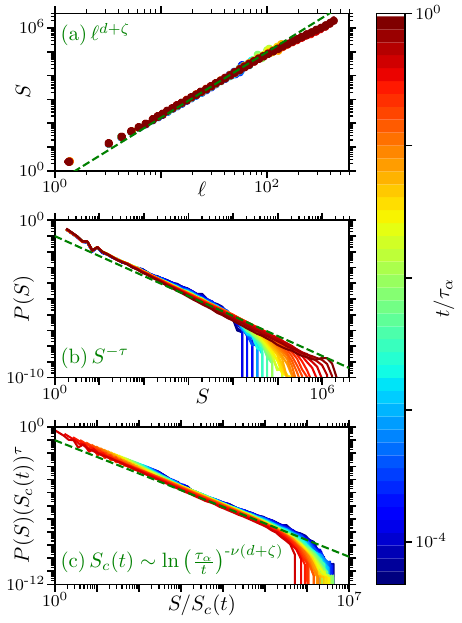}
        \caption{
            Statistics of avalanches at our lowest temperature $T = 0.002$ and biggest system size $L = 2^9 = 512$.
            Each avalanche is defined as a connected cluster in snapshots at different durations $t$ starting from an arbitrary configuration at which the interface is taken flat, see color bar.
            \protect\subref{fig:thermal:fractal}
            The fractal dimension $S \sim \ell^{d + \zeta}$ (binned in $\ell$).
            \protect\subref{fig:thermal:pdf_s}
            The avalanche size distribution $P(S) \sim S^{-\tau}$.
            \protect\subref{fig:thermal:pdf_s_collapse}
            Our prediction in \cref{eq:ellc} together with the fractal dimension collapses the data for $t \leq \tau_\alpha / 2$.
            The depinning predictions (recalled in \cref{tab:depinning}) are indicated by green dashed lines.
        }
        \label{fig:thermal_avalanches}
    \end{figure}

    We measure the avalanche size distribution $P(S)$ and fractal dimension $S \sim \ell^{d + \zeta}$ at our lowest temperature $T = 0.002$, whereby avalanches are defined as connected clusters (see \cref{fig:clusters:clusters}).
    The result is shown in \cref{fig:thermal_avalanches} for different $t$ up to $t = \tau_\alpha$ (see color bar).
    As observed, $S \sim \ell^{d + \zeta}$ for any $t$ as shown in \cref{fig:thermal:fractal}.
    Furthermore, the distribution is consistent with a power law $P(S) \sim S^{-\tau}$ with the depinning prediction for $\tau$ (see main text), up to a cut-off $S_c(t)$ which monotonically increases with $t$ consistent with our prediction in \cref{eq:ellc} as supported by the data collapse in \cref{fig:thermal:pdf_s_collapse}.

    \section{Stability at finite temperature}
    \label{sec:si:tau_alpha}

\begin{figure}[htp]
        \subfloat{\label{fig:thermal:pdf_x}}
        \subfloat{\label{fig:thermal:tau_alpha}}
        \centering
        \includegraphics[width=\linewidth]{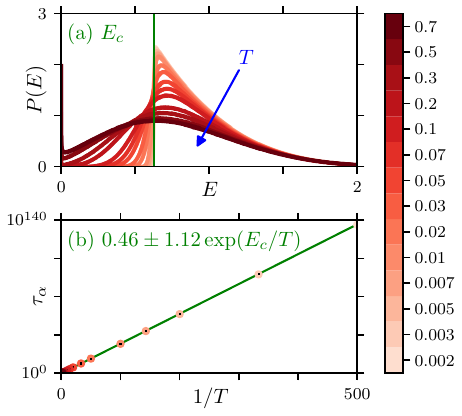}
        \caption{
            \protect\subref{fig:thermal:pdf_x}
            Distribution of activation barriers $P(E)$ at finite temperatures.
            At $T \rightarrow 0$ (and $L \rightarrow \infty$) the distribution displays a gap at $E_c$, as measured in \cref{fig:extremal}.
            This gap corresponds to a characteristic time $\sim \exp(E_c / T)$.
            \protect\subref{fig:thermal:tau_alpha}
            We find that $\tau_\alpha \sim \exp(E_c / T)$ as indicted by the fit in green, whereby $\tau_\alpha$ is defined as the duration at which half of the blocks failed at least once since a random snapshot.
            The values of the temperature $T$ are listed in the color bar.
            All results for our largest system of $L = 2^9$.
        }
        \label{fig:taualpha}
    \end{figure}

    At finite temperature $T$, the dynamics no longer result in an absorbing boundary condition at $E_c$.
    Consequently, the distribution of activation barriers $P(E)$ is no longer gapped at $E_c$, as shown in \cref{fig:thermal:pdf_x}.
    The macroscopic barrier, however, is still set by $E_c$.
    This is shown in \cref{fig:thermal:tau_alpha} whereby we measure $\tau_\alpha$, the time at which half of the blocks yielded at least once, as a function of temperature $T$.
    We indeed find that $\tau_\alpha \sim \exp (E_c / T)$ in \cref{fig:thermal:tau_alpha}.

    \section{Statistics}

    For completeness, we document the extent of our statistics for our largest system of $L^d = (512)^2$ blocks in \cref{tab:statistics}.

    \begin{table}[htp]
        \centering
        \caption{
            Statistics for our largest system of $L^d = (512)^2$ blocks.
            Note that ``sequences'' lists the number of sequences times the number of fails per sequence.
        }
        \label{tab:statistics}
        \begin{tabular}{lrr}
            \toprule
            Temperature & Snapshots & Sequences            \\
            \midrule
            $0^+$       & $391$     & $40 \times 600 L^d$  \\
            $0.002$     & $1262$    & $1149 \times 20 L^d$ \\
            $0.003$     & $1102$    & $992 \times 20 L^d$  \\
            $0.005$     & $1231$    & $1129 \times 20 L^d$ \\
            $0.007$     & $1125$    & $1017 \times 20 L^d$ \\
            $0.01$      & $1167$    & $1057 \times 20 L^d$ \\
            $0.02$      & $1232$    & $1124 \times 20 L^d$ \\
            $0.03$      & $1235$    & $1123 \times 20 L^d$ \\
            $0.05$      & $1132$    & $1018 \times 20 L^d$ \\
            $0.07$      & $1125$    & $1005 \times 20 L^d$ \\
            $0.1$       & $287$     & $241 \times 20 L^d$  \\
            $0.2$       & $399$     & $333 \times 20 L^d$  \\
            $0.3$       & $214$     & $167 \times 20 L^d$  \\
            $0.5$       & $275$     & $224 \times 20 L^d$  \\
            $0.7$       & $220$     & $179 \times 20 L^d$  \\
            \bottomrule
        \end{tabular}
    \end{table}

\end{appendices}

\end{document}